\documentclass[letterpaper,dvipsnames]{article}
\pdfoutput=1 

\usepackage{jheppub} 
\usepackage{hyperref}

\usepackage{amsmath}
\usepackage[T1]{fontenc} 
\usepackage{braket}
\usepackage{subcaption}
\usepackage{soul}
\usepackage{array}
\usepackage{xcolor}
\usepackage{cancel}
\usepackage{slashed}
\usepackage{comment}
\usepackage[many]{tcolorbox}    	

\newtcolorbox{boxA}{
    fontupper = \bf,
    boxrule = 1.5pt,
    colframe = black 
}

\newtcolorbox{boxE}{
    enhanced, 
    boxrule = 0pt, 
    borderline = {0.75pt}{0pt}{main}, 
    borderline = {0.75pt}{2pt}{sub} 
}

\newcommand{\Mpl}{M_{\rm pl}}
\newcommand{\hc}{\rm h.c.}

\newcommand{\PQ}{{\sf PQ} }

\newcommand{\Calmbox}{\includegraphics[width=6pt]{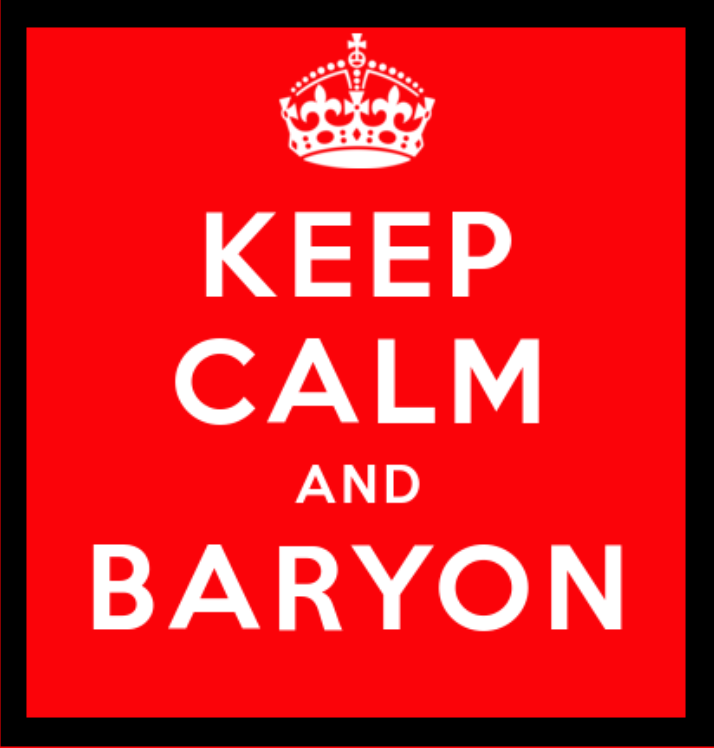}}

\newcommand{\klr}{K\"{a}hler }

\usepackage[compat=1.1.0]{tikz-feynman}

\author[a,d]{Prateek Agrawal,}
\author[b]{Anson Hook,}
\author[a]{Vazha Loladze,}
\author[a,c]{and Mario Reig}

\affiliation[a]{Rudolf Peierls Centre for Theoretical Physics, 
University of Oxford, Parks Road, Oxford OX1 3PU, United Kingdom}

\affiliation[b]{Maryland Center for Fundamental Physics, Department of Physics, University of Maryland, College Park, MD 20742, U.S.A.}

\affiliation[c]{Theoretical Physics Department, CERN, 1211 Geneva 23, Switzerland}

\affiliation[d]{Department of Physics, University of California, Santa Barbara, CA 93106, USA}

\title{Axion Quality Problem: Keep Calm and Baryon}

\abstract{
Axion models 
generically suffer from a severe quality problem when coupled to gravity. 
In this article we provide a very simple model with a high quality axion. The axion is a pseudo-Nambu-Goldstone boson of the baryon number symmetry, $U(1)_B$, of a new composite sector that breaks $U(1)_B$ spontaneously when it confines. 
A controlled example is a supersymmetric QCD (SQCD) with $N_c = N_f$. The axion shift symmetry is automatically protected due to the high dimension of the gauge-invariant baryon operator, with the Peccei-Quinn breaking operators arising at dimension $N_c+2$. The standard model gauge group is embedded as a subgroup of the flavor symmetry group of SQCD that has an anomaly with $U(1)_B$, generating the standard coupling with gluons.}

\begin{document} 

\maketitle

\section{Introduction}

The QCD axion~\cite{Peccei:1977hh,Weinberg:1977ma,Wilczek:1977pj} is one of the best motivated particles beyond the Standard Model (BSM). Its ability to solve the strong CP problem, explaining the absence of the neutron electric dipole moment, as well as the dark matter puzzle~\cite{Preskill:1982cy,Abbott:1982af,Dine:1982ah}, has led to a large experimental program trying to find this particle~\cite{Graham:2015ouw,Irastorza:2018dyq,OHare:2024nmr}. The axion arises as the pseudo-Nambu-Goldstone boson of a spontaneously broken Peccei-Quinn (PQ) symmetry $U(1)_{\PQ}$, a global symmetry that has a mixed anomaly with $SU(3)_c$~\cite{Dine:1981rt,Zhitnitsky:1980tq,Kim:1979if,Shifman:1979if}. 
The stringent limits on the neutron EDM~\cite{Abel:2020pzs} require an abnormally pristine PQ symmetry, the so-called axion quality problem~\cite{Georgi:1981pu,Kamionkowski:1992mf,Barr:1992qq,Holman:1992us}. In view of the expectation that quantum gravity effects break every global symmetry, this becomes a serious problem.

The quality of the shift symmetry can be exponentially good in higher-dimensional constructions~\cite{Choi:2003wr} and in string theory~\cite{Svrcek:2006yi,Arvanitaki:2009fg}. In this case, the axion shift symmetry descends from a higher-form global symmetry that requires a non-local effect to break it, which can generically be exponentially small.
A rationale for a high-quality shift symmetry for the QCD axion is harder to obtain within a four-dimensional field theory construction. The key problem is that local operators that break the PQ symmetry need to be highly suppressed, but the cutoff of the theory does not respect any global symmetries even approximately (if it did then the quality problem is deferred to the UV theory). Within the logic of 4d effective field theory, the only suppression available is a high dimensionality of allowed PQ-violating operators, which has to be imposed by hand. For example, in simple models where the axion is the phase of a complex scalar field $\Phi$,  operators $\mathcal{O}\sim c\,{\Phi^N}/{\Mpl^{N-4}}$ with $N < 12$ (for $f_a \sim 10^{11}$ GeV, and $c=O(1)$) need to be disallowed. This can be justified, for example, by imposing a $\mathbb{Z}_{12}$ discrete gauge symmetry.
 
Composite axion models~\cite{Kim:1984pt,Choi:1985cb} are attractive because they do not introduce new fundamental scalars.    
The axion is the pseudo-Nambu-Goldstone boson (pNGB) associated with the spontaneous breaking of a flavor symmetry, similar to pions in QCD. These models also suffer from the quality problem, but can be imbued with additional structure so that the PQ symmetry becomes accidental~\cite{Randall:1992ut}. 
These constructions have received a lot of attention in recent years in different forms~\cite{Redi:2016esr,Lillard:2017cwx,Gavela:2018paw,Cox:2019rro,Ardu:2020qmo,Yin:2020dfn,Contino:2021ayn,Cox:2023dou,Nakagawa:2024kcb,Gherghetta:2025fip,Gherghetta:2025kff}, but can involve complicated additional structure that undermines the simplicity of the original composite axion. See also  \cite{Babu:2002ic,Lee:2011dya,Harigaya:2013vja,Duerr:2017amf,Bhattiprolu:2021rrj} for other realisations where the PQ symmetry arises accidentally due to gauge symmetries.

In this paper we provide a very simple incarnation of a high quality PQ symmetry -- it is the baryon number symmetry of a new confining sector. A controlled implementation of the idea is found in
SQCD~\cite{Seiberg:1994bz,Intriligator:1994sm} with $N_c = N_f$. This theory is known to have a moduli space with massless mesons and baryons below the confining scale. Remarkably, once the effects of explicit global symmetry breaking are included, the theory generically picks a vacuum where the baryons take a vacuum expectation value (vev), $\langle B\rangle \neq 0$, and baryon number $U(1)_B$ is broken spontaneously. Thus, a high-quality pNGB is a generic prediction of SQCD with $N_c = N_f$! Unlike in pion-like composite axion models, baryon number violation decreases exponentially with $N_c$. After including SUSY breaking terms, the leading operator breaking the PQ symmetry appears at dimension $N_c+2$. A similar idea was employed in Ref.~\cite{Baumann:2010nu} to get a flat potential for inflation.

The PQ mechanism requires the presence of a spontaneously broken symmetry which has a mixed anomaly with QCD. To introduce the  anomaly $[SU(3)_c]^2\times U(1)_B$, we embed QCD in a subgroup of the flavor symmetry, $SU(N_f)_L$, which has a mixed `t Hooft anomaly $[SU(N_f)_L]^2\times U(1)_B$. This gives an Adler-Bell-Jackiw (ABJ) anomaly, providing the right ingredients for implementing the PQ solution to the strong CP problem. One simple possibility is a theory with $N_c = N_f = 10$ where we gauge an $SO(10)$ subgroup of the flavor symmetry, $SU(10)_L$. The SM fermions are $SU(N_c)$ singlets that transform as $\mathbf{16}$ of the gauged $SO(10)$.

A construction where the PQ symmetry is related to  baryon number in SQCD was presented by Lillard and Tait~\cite{Lillard:2018fdt} with slightly different aims. 
In their work, they consider an extended model with 2 confining groups and fermions transforming as bifundamental representations, addressing the axion quality problem, the $\mu$-problem and include an additional gauge sector connected to $U(1)_{B-L}$. In our work, we are interested in identifying the minimal structure required to solve the axion quality problem and find, remarkably, that no structure beyond the SQCD theory is needed.

\section{Supersymmetric QCD}
Supersymmetric QCD is a very well-studied generalization of QCD with $\mathcal{N}=1$ supersymmetry~\cite{Seiberg:1994bz,Intriligator:1994sm}. It is a gauge theory with a gauge group $SU(N_c)$ and $N_f$ chiral superfields (quarks, $Q$) in the fundamental representation, and $N_f$ chiral superfields (anti-quarks, $\tilde{Q})$ in the anti-fundamental representation. The flavor symmetry of the model is  $SU(N_f)_L\times SU(N_f)_R\times U(1)_B$. In this paper will focus on the case $N_f = N_c = N$.

The moduli space of SQCD is well-understood. The moduli space for the theory can be written in terms of the ($SU(N_c)$) gauge-invariant superfields, the \emph{baryons} ($B,\tilde{B}$) and the \emph{mesons} $M^i_j$, given in terms of the quarks as 
\begin{align}
  M^{i}_{\ j} = Q^i \tilde{Q}_j  \,, 
  \quad B = \epsilon_{i_1\ldots i_{N_c}} Q^{i_1}\ldots Q^{i_{N_c}} \,, 
   \quad \tilde{B} = \epsilon_{i_1\ldots i_{N_c}} \tilde{Q}^{i_1}\ldots \tilde{Q}^{i_{N_c}}
   \,.
\end{align}
where $i,j$ are flavor indices. The moduli space is modified by strong dynamics~\cite{Seiberg:1994bz} to satisfy the constraint, 
\begin{align}
\det M - B \tilde{B} = \Lambda^{2N_c} \,,
\end{align}
where the RHS is a quantum effect from confinement of the theory at the scale $\Lambda$~\cite{Seiberg:1994bz,Intriligator:1994sm}. 
At large field values, this constraint can be seen classically from the definition of $M$ and $B$ in terms of $Q,\tilde{Q}$. Quantum effects smooth out the singular points on the moduli space.

There are two loci of enhanced symmetry on the moduli space, where $B=\tilde{B}=0$, or $\det M = 0$.
For our application, we will be interested in the latter, where confinement breaks baryon number spontaneously.

\section{Our model}

In our application, the $U(1)_B$ will play the role of a high quality Peccei-Quinn (PQ) symmetry for the axion.
To solve the strong CP problem, this symmetry should have an ABJ anomaly with the standard model QCD. A simple way to achieve this is to weakly gauge a subgroup of the $SU(N_f)_L$ flavor symmetry by standard model color, $SU(3)_c$. The mixed `t Hooft anomaly in the symmetry  $[SU(N_f)_L]^2\times U(1){_B}$ implies that upon gauging, we get the desired PQ anomaly $[SU(3)_c]^2\times U(1){_\PQ}$.
The spontaneous breaking of the baryon number after confinement will produce the axion which couples to the gluons with the standard QCD axion coupling.

There are many possibilities to embed the SM into $\mathcal{G}_f\subset SU(N_f)_L$ group. The restriction for the mechanism to work is that the mixed anomaly $[SU(3)_c]^2\times U(1)_B$ is non-zero while the anomaly $[SU(3)_c]^3$ is zero. For example, we can choose 6 SQCD quarks $Q$ to transform in a $3 \oplus \bar{3}$ representation of $SU(3)_c$. 
An elegant way to embed the SM gauge group in $SU(N_f)_L$ while avoiding gauge anomalies is to model the SM as an $SO(10)$ unified theory, so that for $N_c=N_f = 10$, $Q$ transforms in the fundamental representation of $SO(10)$. 
The field content of the model is given in Table~\ref{table:content}. 

It is also possible to consider models with a larger confining group, $N_c = N_f > 10$ and gauge a subgroup of $SU(N_f)_L$ by the SM gauge group. However, if there are quarks which do not carry any weakly gauged flavor quantum numbers, they will likely pair up with the anti-fundamentals $\widetilde{Q}$ to get a mass at the Planck scale, and decouple. It is easy to build models which gauge e.g.~an $SO(10)\times SO(N_f - 10)$ subgroup of $SU(N_f)_L$ that avoid this problem. We will not pursue this avenue in detail in this paper.
We will keep the discussion general in the following assuming a general $N_c = N_f \geq 10$ confining SQCD theory, with the example with $N_c = N_f = 10$ as the prototype.

\begin{table}[h]
\centering
\begin{tabular}{|c|c||c|c|c|c|}
\hline
 Symmetry & \multicolumn{1}{c||}{Gauge}&\multicolumn{4}{c|}{Global}\\
\cline{1-6}
 & $SU(N)$ & $SU(N)_L$& $SU(N)_R$ & $U(1)_{B}$ & $U(1)_R$\\
& & ({\footnotesize $\supset {\rm gauged}\ SO(10)_{\rm SM}$})  & & & \\
\hline
$Q$ & \Calmbox & \Calmbox (\Calmbox) & $1$ & $+1/N_c$ & $0$ \\
$\tilde{Q}$ & $\overline{\Calmbox}$ &  $1$ ($1$) & $\overline{\Calmbox}$ & $-1/N_c$ & $0$ \\

\hline
\end{tabular}
\caption{Chiral superfields in SQCD with gauge group $SU(N_c)$ and $N_f$ flavors with $N_c = N_f=N$.  The SM is contained in an $SO(10)$ which is embedded in $SU(N)_L$, weakly gauging the flavor symmetry.}
\label{table:content}
\end{table}

In any case, the SM fermions will be $SU(N_c)$ singlets. PQ breaking, and therefore the axion decay constant, is related to the scale of spontaneous breaking of baryon number.
In the case described in Table~\ref{table:content}, the anomaly coefficient is easily calculated in terms of the Dynkin index, $T_{10}=1$, the number of $SO(10)$ fundamentals and their $U(1)_B$ charge, $\mathcal{A}=N_c\times 1\times 1/N_c=1$. This indicates that this class of composite axion models does not suffer from the cosmological domain wall problem.

\begin{figure}[t]
\centering
\includegraphics[width=.65\linewidth]{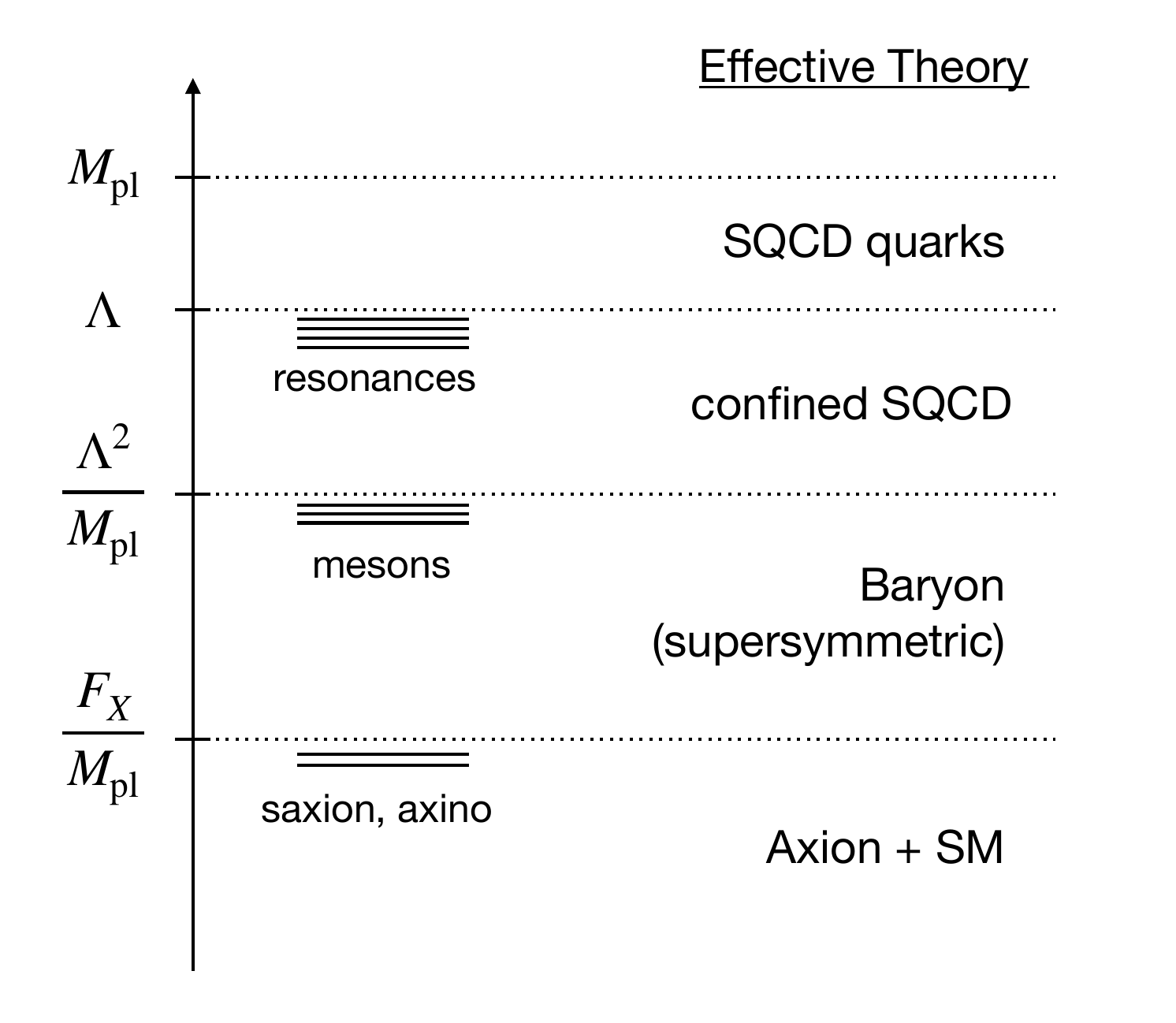}
\caption{A cartoon of the relevant energy scales and the different phases of the $SU(N_c)$ theory. Below the scale $F_X/\Mpl$ only the axion survives as a massless degree of freedom up to effects from QCD and higher-dimensional operators breaking its shift-symmetry.
} 
\label{Fig: fig}
\end{figure}

Our model deforms the standard SQCD theory in 3 ways: by gauging part of the flavor symmetry, by allowing higher-dimensional operators, and by including supersymmetry breaking effects.  A simple picture of the energy scales in our theory and what occurs at each scale is shown in figure~\ref{Fig: fig}.  In the UV there is the theory of weakly interacting quarks.  At the confinement scale $\Lambda$, the quarks get confined and it becomes a theory of massless mesons and baryons.  Eventually Planck scale-suppressed higher-dimensional operators give a mass to the mesons and only the baryons (with the axion being the phase) remain.  Finally, SUSY breaking gives a mass to the saxion and the axino.

\subsection{Weak coupling quark picture}

The quality of the axion in this framework is determined by symmetry breaking Planck suppressed operators.  The higher-dimensional operators induced by gravity will, in general, break all global symmetries.
The most important higher-dimensional operators will be ones that break the global $U(1)_B$ and $SU(N_f)_R$ symmetries of SQCD,
\begin{align}
\label{eq:deformed_superpotential}
W &= 
\frac{c_{jk}}{\Mpl} (Q_i \widetilde{Q}_j) (Q_i \widetilde{Q}_k)
+ \frac{\lambda}{\Mpl^{N-3}} 
Q_{[i_1}\ldots Q_{i_N]}
+ \frac{\widetilde\lambda}{\Mpl^{N-3}}
\widetilde{Q}_{[i_1}\ldots \widetilde{Q}_{i_N]}\,,
\end{align}
where indices in square brackets indicate anti-symmetrization. The coefficients $\lambda$, $\tilde{\lambda}$, and $c_{jk}$ are order one.
A mass term for the quarks is not generated due to the gauged $SO(10) \subset SU(N_f)_L$ symmetry. When $N_f > 10$ the structure of the leading operators remains the same if all quarks $Q$ have weakly gauged flavor quantum numbers.

Aside from internal symmetry breaking, SUSY breaking will also be important.  
For simplicity, we take the framework of gravity mediation, where SUSY breaking is carried by a chiral superfield $X = F_X \theta^2$ which couples with $\Mpl$ suppressed operators. We will assume that the SUSY-breaking effects are smaller than the confinement scale, so that we can treat confinement supersymmetrically. We study effects of SUSY breaking directly in the confined theory, using naive dimensional analysis to estimate the size of various effects.

\subsection{Meson/Baryon picture}
Below the confinement scale, the low-energy effective theory is represented in terms of the baryons $(B,\tilde{B})$ and mesons $M$ superfields.  Supersymmetry makes it possible to analytically track confinement and calculate the generated superpotential in the IR.  While the moduli space is described by baryons and mesons, the moduli space is constrained and not all of them are present at the same time in the IR effective field theory.  
It is well-known that
in the theory with $N_c=N_f$
confinement modifies the constraint on the moduli space~\cite{Seiberg:1994bz},
\begin{align}
 \frac{1}{\Lambda^{N-2}}\det M - B \tilde{B} - \Lambda^{2} &=0
 \,.
\end{align}
We have normalized the baryon and meson fields to have mass dimension one. 
Because we are interested in the vacuum where $\det M = 0$ and $B \tilde{B} = -\Lambda^2$, we can use the constraint to remove e.g.~$\widetilde{B}$ from the theory.
At this stage the massless fields are the mesons, which match the 't Hooft anomalies for the chiral symmetry, and the baryon $B$, which contains the axion, saxion and axino. This field is massless due to a combination of supersymmetry and the axion's pseudo-Goldstone boson nature.

Global symmetry and supersymmetry breaking effects will eventually lift all the flat directions. There are three important effects: chiral symmetry breaking operators, supersymmetry breaking operators, and baryon number breaking operators. The chiral symmetry breaking operators arise at dimension-4 in the superpotential (equation~\eqref{eq:deformed_superpotential}), which will generate a mass term for the mesons in the superpotential of the order $\Lambda^2/\Mpl$. We assume that this is the largest breaking effect, and that the supersymmetry breaking effects (parametrized by $F_X$) are smaller than this scale, see figure~\ref{Fig: fig}. 

We can integrate out the mesons supersymmetrically,
\begin{align}\label{eq:meson_mass}
W \supset c_{jk}\frac{\Lambda^2}{\Mpl} M^j_{\,\,i} M^i_{\,\,k}
\,,
\end{align}
which generically stabilizes the moduli at $M^j_{\,\,i}=0$.
Below the scale $\Lambda^2/\Mpl$, we are left with a simple supersymmetric theory with only one chiral superfield, the baryon $B$, which contains the pNGB of baryon number breaking. 
In the supersymmetric limit, the leading \klr potential and superpotential are
\begin{align}
K &= B^\dagger B+\tilde{B}^\dagger\tilde{B} = B^\dagger B
 +
 \left|
 \Lambda^2 B^{-1}
 \right|^2\,,\\
W &\sim \frac{\Lambda^{N-1}}{\Mpl^{N-3}} B +  \frac{\Lambda^{N+1}}{\Mpl^{N-3}} \frac{1}{B} 
\,,
\end{align}
where are leaving $O(1)$ dimensionless coefficients of operators above implicit. The baryon number breaking superpotential comes from the high dimensional operator in equation~\eqref{eq:deformed_superpotential}. This is an extremely high dimensional operator, and SUSY breaking terms will be much more relevant in giving particles a mass. There is also a Wess-Zumino-Witten term between the SM gauge bosons and the baryon field to match anomalies. We will not need its detailed supersymmetric form and include it below the scale of SUSY breaking.

For gravity mediated SUSY breaking, a simple way to obtain all supersymmetry breaking operators is to elevate all couplings to real superfields with non-zero $\theta^2, \theta^2 \bar{\theta}^2$ components coming from the SUSY breaking superfield $X$. For the saxion, this gives us 
\begin{align}
L \supset \int d^4 \theta
\frac{X^\dagger X}{\Mpl^2} B^\dagger B  + \frac{X^\dagger X}{\Mpl^2}
\frac{\Lambda^4}{B^\dagger B} \sim \frac{F_X^2}{\Mpl^2} B^\dagger B  + \frac{F_X^2}{\Mpl^2}
\frac{\Lambda^4}{B^\dagger B}\,,
\end{align}
which gives a vev $\langle B^\dagger B \rangle \sim \Lambda^2 \equiv f^2$. 

Higher order terms in the \klr potential (either through series expansion of the $1/B$ terms or otherwise) give a Majorana mass for the axino after SUSY and the spontaneous $U(1)_B$ breaking, for instance,
\begin{align}
    L
    \supset
    \int d^4 \theta 
    \frac{X+X^\dagger}{\Mpl} \frac{1}{\Lambda^2} (B^\dagger B)^2
    \sim \frac{F_X}{\Mpl} \frac{f^2}{\Lambda^2} \psi_B \psi_B
    \,.
\end{align}

Writing $B = (f+s)\exp({i a})$, below the scale $F_X / \Mpl$ the saxion and axino gain a mass and we get only the axion,
\begin{align}
\mathcal{L}
&=
\left(f^2 + \frac{\Lambda^4}{f^2}
\right)
(\partial a)^2 + \frac{g^2}{32 \pi^2} a G \tilde G
\, ,
\end{align}
where we have now explicitly included the axion coupling to QCD coming from the WZW term.

\subsection{Axion quality}

Baryon number-violating deformations to the superpotential induce small shift-symmetry-breaking effects for the QCD axion. The allowed explicit breaking effects are small, however, providing a candidate for a high-quality axion. In this subsection, we quantify these effects more precisely.

The baryon number breaking effects induced by gravity will generate a potential for the axion. By matching with equation~\eqref{eq:deformed_superpotential} the leading operator in terms of the superfield $B$ is,
\begin{align}
    W
    \supset
    \lambda \frac{\Lambda^{N-1}}{\Mpl^{N-3}} B \,.
\end{align}
The leading contribution to the axion potential comes from the SUSY breaking contribution (as long as the hierarchy between $F_X$ and $\Lambda$ is not too large). Again, the SUSY breaking can be included by giving $\lambda$ an $F$-term. This generates a potential for the axion,
\begin{align}\label{eq:V_BV_potential}
    V_{\rm BV}
    &= \int d^2 \theta
    \frac{X}{\Mpl} 
    \frac{\Lambda^{N-1}}{\Mpl^{N-3}} B + {\hc}
    =
    F_X
    \frac{\Lambda^{N-1}}
    {\Mpl^{N-2}} f \cos (a  + \delta)\,,
\end{align}
where we have used $\lambda|_{\theta^2} =F_X/\Mpl$. 

The axion potential above, $V_{\rm BV}$, has in general an $O(1)$ phase offset, $\delta$, with respect to the axion potential from QCD effects. As usual, the induced shift with respect to the CP conserving minimum is
\begin{align}\label{eq:eff_theta}
    \theta_{\rm eff}=\frac{\Lambda^{\prime\,4}\sin\delta}{f_\pi^2 m_\pi^2+\Lambda^{\prime\,4}\cos\delta}
    \sim
    \frac{F_X \Lambda^N}{\Mpl^{N-2}f_\pi^2 m_\pi^2  }
    \,.
\end{align}
Here $\Lambda^{\prime\,4}=F_X
    \frac{\Lambda^{N-1}}
    {\Mpl^{N-2}} f$ corresponds to the scale of the axion potential induced by the baryon number-violating deformation.
    
Solving the strong CP problem requires $\theta_{\rm eff}\lesssim 10^{-10}$~\cite{Abel:2020pzs}. To be more quantitative, let us consider the benchmark model with $N_c=10$, a decay constant around $\Lambda \sim f \sim 10^{11}$ GeV, and $F_X\lesssim (10^{11}$ GeV$)^2$. In this case the amplitude of the potential due to the baryon-number-violating deformation is $|V_{\rm BV}|\lesssim 10^{-15}\text{ GeV}^4$. 
This leads to a misalignment with respect to the CP preserving vacuum around $\theta_{\rm eff}\lesssim 10^{-11}$, which suffices to solve the strong CP problem. Larger decay constants, however, require a larger number of colors or smaller supersymmetry breaking scale in order to maintain $\theta_{\rm eff}$ small.

\begin{figure}[t]
\centering
\includegraphics[width=.7\linewidth]{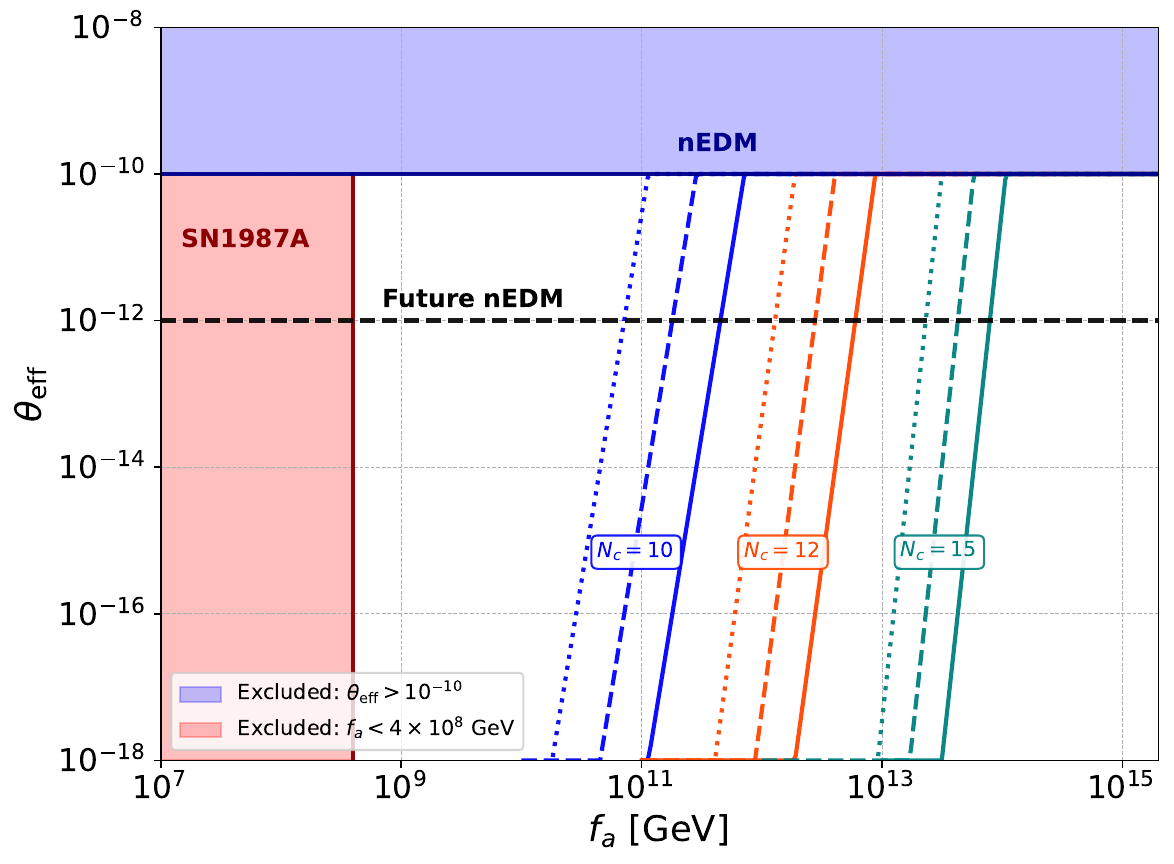}
\caption{Effective theta angle $\theta_{\rm eff}$ as a function of the decay constant for theories with different number of colors, $N_c$, in \textcolor{blue}{blue}, \textcolor{RedOrange}{orange red} and \textcolor{teal}{teal}. Dotted, dashed and solid lines correspond to different SUSY breaking scales, $\sqrt{F_X}=10^{11}, 10^9,10^7$ GeV, respectively, for each $N_c$. More choices for $\sqrt{F_X}$ and $N_c$ are in principle allowed, with the predicted $\theta_{\rm eff}$ changing accordingly. For reference, the lower bound to $f_a$ from astrophysics~\cite{Carenza:2019pxu,Buschmann:2021juv,Springmann:2024ret,Fiorillo:2025gnd}  (see~\cite{Caputo:2024oqc} for a recent review), as well as upper bound to $\theta_{\rm eff}$~\cite{Abel:2020pzs} are shown in shaded \textcolor{BrickRed}{red} and \textcolor{Blue}{blue}, respectively.}
\label{Fig:fig_main}
\end{figure}

The supersymmetry breaking scale $F_X$ affects the parameter space weakly. In figure~\ref{Fig:fig_main} we show the parameter space where the QCD axion has sufficient quality to solve the strong CP problem. In parts of the parameter space the predicted $\theta_{\rm eff}$ may be detectable in future nEDM experiments, which will improve the current bound by around 2 orders of magnitude~\cite{pEDM:2022ytu,n2EDM:2021yah,TUCAN:2022koi,Alarcon:2022ero}.

\subsection{Meson masses and Landau pole}

The presence of light meson superfields with SM charge modifies the beta function of the SM gauge group factors, possibly introducing Landau poles below the confinement scale. To illustrate the issue, let us take the example of a theory with $N_c=N_f=10$ and study the running of the QCD gauge coupling with a MSSM-like spectrum. The mesons transform in the fundamental representation of the weakly gauged $SO(10)_{\rm SM}$, which under the SM gauge group have quantum numbers
\begin{equation}
    \pi\sim (\mathbf{3},1,-1/3)+(\bar{\mathbf{3}},1,1/3)+(1,\mathbf{2},-1/2)+(1,\mathbf{2},1/2)\,.
\end{equation}
These modify the beta function of QCD, which becomes
\begin{equation}
    b_{\rm sqcd}= 2 N_g-9+N_{\pi} \,.
\end{equation}
Here $N_g=3$ is the number of MSSM generations and $N_{\pi}=10$ the number of $SO(10)_{\rm SM}$ fundamentals. 

For simplicity, we assume that the mass scale of colored superpartners is kept fixed around $m_{\rm soft} = 1$ TeV. We can use the SM running up to $m_{\rm soft}$, which yields $\alpha_s(m_{\rm soft})=0.09$. To get the coupling at high scales, we use MSSM running up to the meson mass scale $m_M$, and then include the meson contribution in the running above their mass,
\begin{align}
\label{eq:RGE}
    \alpha_s^{-1}(\mu) =  \alpha_s^{-1}(m_{\rm soft}) 
    +\frac{b_{\rm mssm}}{2\pi}
    \log \frac{m_{\rm soft}}{m_M} 
    +\frac{b_{\rm sqcd}}{2\pi}
    \log \frac{m_M}{\mu}\,.
\end{align}
which yields the expression for the Landau pole ($\alpha^{-1}(\Lambda_{\rm LP})\rightarrow 0$), 
\begin{align}
\label{eq:RGE2}
    \Lambda_{\rm LP}
    = m_M 
    \exp\left(
    \frac{2\pi}{\alpha_s(m_{\rm soft}) b_{\rm sqcd}} 
    \right)
    \left(
    \frac{m_{\rm soft}}{m_M}
    \right)^{\frac{b_{\rm mssm}}{b_{\rm sqcd}}}
    \simeq
    2\times10^7{\ \rm GeV}
     \left(
    \frac{m_M}{\rm 1\ TeV}
    \right)^{10/7}.
\end{align}
For a confinement scale $\Lambda\sim \mathcal{O}(10^{11})$ GeV, if the meson mass comes from the Planck suppressed operator in Eq.~\eqref{eq:meson_mass}, it lies around $\mathcal{O}(10)$ TeV. In this case we find a Landau pole at around $\Lambda_{\rm LP}\sim 10^8$ GeV, indicating that QCD becomes non-perturbative before the $SU(N)$ confinement scale, $\Lambda$, which indicates that our description using weakly gauged flavour symmetry is not consistent. We explore possible ways that allow $\Lambda_{\rm LP}\gtrsim \Lambda$. 


The problem follows from light meson masses, which reflect the fact that the chiral symmetry in the SQCD sector is broken only at the Planck scale. If there are other sources of chiral symmetry breaking at lower scales, the meson masses can be made much larger without affecting the rest of the mechanism. A simple way to do this is to add new chiral superfield transforming as $\phi \sim (\mathbf{10},\mathbf{10})$ under the flavour group and add to the superpotential 
\begin{align} 
W_\phi = y_{jk} \phi_{ij} Q_i \tilde Q_j + M_\phi \phi^2
\end{align}
These operators break the flavour symmetry $SU(10)_R$ explicitly. Integrating out $\phi$ leads to the effective operator,
\begin{equation}
W \supset \frac{c_{jk}}{M_\phi}(Q_i\tilde Q_j)(Q_i\tilde Q_k)\,.
\end{equation}
which induces a meson mass suppressed only by the $\phi$ mass, $m_M\sim\frac{\Lambda^2}{M_\phi}$. The meson mass can be easily be as high as O($\Lambda$), which moves the Landau pole beyond the GUT scale. The quality of the meson chiral symmetry does not interfere with the quality of the baryon symmetry, and therefore does not affect our mechanism.

For $N_c > 10$, a similar mechanism will continue to work. However, the problem is somewhat less severe in this case. As shown in Fig.~\ref{Fig:fig_main}, a larger number of colours $N_c$ larger allows to make $\Lambda$ larger while being consistent with axion quality. This in turn increases the meson mass quadratically and the Landau pole occurs at higher energies. 

A full model will include the details of grand unification (if it indeed is a GUT model), GUT breaking, and running above the GUT scale. The details are highly dependent on the model implementation, and are beyond the scope of the current work which focuses on the simple mechanism for a high quality axion.

\section{Conclusion}
In this article we have explored a very simple construction where the QCD axion arises from spontaneous breaking of baryon number in a $SU(N_c)$ SQCD theory. As shown in equations~\eqref{eq:V_BV_potential} and \eqref{eq:eff_theta}, the quality grows rapidly with increasing number of colors due to compositeness of baryons. 

While the number of colors is arbitrary (as long as the flavor symmetry is large enough to accommodate QCD), a compelling, minimal theory arises for $N_c=N_f=10$, where the SM is embedded as the $SO(10)$ subgroup of the $SU(10)_L$ flavor symmetry. In this case, compatibility with standard lower bounds on the axion decay constant and maintaining the solution to the strong CP problem decay constant leads to the range $10^8  \text{ GeV}\lesssim f_a \lesssim 5\times 10^{11} \text{ GeV}$. 

The construction presented here opens new possibilities for composite axion model building.  
Interesting questions that remain include 
studying supersymmetry breaking in more detail, for example how it is communicated to the visible sector and how this could affect the meson as well as baryon masses is also an important open question that we leave for future work.
It will be nice to study the interplay of the mechanism discussed here with high quality extra-dimensional axions, e.g.~in the context of holography.

\acknowledgments
VL and MR would like to thank Deutsche Bahn for an ``unexpected'' delay that allowed us to try the Tunnel Doner and start the first version of this manuscript. We thank Munich Institute for Astro-, Particle and BioPhysics (MIAPbP) for support and hospitality funded by the Deutsche Forschungsgemeinschaft (DFG) Excellence Strategy – EXC-2094 – 390783311 and the organisers of the workshop Navigating New Horizons, where this work was started. PA and
VL are supported by the STFC under Grant No.~ST/X000761/1.
AH is supported by NSF grant PHY-2514660 and the Maryland Center for Fundamental Physics.

\bibliographystyle{JHEP}

\providecommand{\href}[2]{#2}\begingroup\raggedright\endgroup

\end{document}